Vibe Coding, Interface flattening

**H. Kevin Jin**

University of Amsterdam

kevin.jin@student.uva.nl

**Table of Contents**

## 1. Introduction

The emergence of large language models has precipitated what may be understood as a groundbreaking reconfiguration in the materialist foundations of human-computer interaction. Among the various applications of LLM-adjacent technologies, "vibe coding", a paradigm of software development characterised by natural language interaction with AI toolchains represents perhaps the most significant departure from established modes of human-computer interaction. This paradigm operates as a qualitatively distinct form of computational mediation that challenges prevailing assumptions about programming practices and human-computer interaction. What interests us in vibe coding is not precisely its boasted enhancements in developer efficiency or accessibility, but rather how it mobilises a technological constellation of remote computation, knowledge aggregation, and probabilistic inference that *alters* the mediating structures through which human intent becomes computational operation. This reconfiguration suggests what may be characterised as *interface flattening*: a process wherein previously distinct interface modalities converge into a singular, model-mediated substrate that obscures traditional boundaries between graphic user interfaces (GUI), command-line environments (CLI), and application programming interfaces (API). *Interface flattening* operates through two interrelated mechanisms. First, it depends on a *computational configuration* optimised for *models* that is materially and spatially distinct from other computer systems. This substrate re-layers existing interfaces and introducing a probabilistic mediator to human-computer interactions. Second, it produces apparent integration of previously distinct interface modalities, creating a *meta-interface* that affords natural language interactions and ostensibly offsets specialised technical knowledge.

The selection of vibe coding as an analytical case arise from its kinship with the programming practice. Programming has historically constituted a specialised practice requiring mastery of domain-specific interfaces. Yet this specialisation obscures how programming practice *emerged from* and remains *connected to* broader histories of computer use. True that programming invokes particular forms of symbolism and abstraction (syntax, proto-cols, APIs, console metrics) that differ from audiovisual and textual cues characterising "everyday" user interfaces, yet these differences arise from the same co-constructed social-technical background where programming is neither different nor more privileged than other forms of human-computer interaction in *creating effects*. Analysing the vibe coding paradigm with a *historical sensitivity* then offers a proposal for understanding how emergent AI technologies reconfigure the relationship between human intention and computational operation, where *supposedly essential* requirements for computational control become now bypassable through natural language interaction with AI/LLM toolchains.

This paper proceeds through three movements. Following this introduction, I examine the theoretical foundations necessary for understanding programming practices as historically specific forms of human-computer interface, drawing primarily upon Friedrich Kittler's materialist analysis of computational media and Alexander Galloway's work on interface theory and protocol control. This theoretical framework establishes programming as a locally stabilised mode of translating human intent into machinic effect, subject to ongoing reconfiguration through socio-technical shifts. I then turn to an analysis of vibe coding as an exemplary case of interface flattening, examining both its empirical manifestations and its theoretical implications for understanding emergent AI technologies. I conclude by considering how interface flattening may offer new directions for critical engagement with the political and material dimensions of AI-mediated human-technology interactions.

## 2. Programming/Interface

For media scholars, what does it mean to study vibe coding, or any coding at all? On one hand, its significance is secured by the infrastructural role of computing hardware and code operation in contemporary media systems. On the other hand, precisely because computation undergirds mediated reality, the coding practice and its material constellations offer a concrete site for examining how human intention is translated into machinic effect, and how technical intricacies become legible as "code." To understand programming is, therefore, to examine the *interface conditions* under which the human conveys intent to the technical, and how the technical is rendered as a field of action for the human. Two thinkers offer a productive frame for modelling how code, machines, and people meet in *practice*. Kittler's insistence that "there is no software" compels an analysis beneath the symbolic layer of code to machine operations of voltage differences (Kittler 1995; 1997, 150). From this perspective, the user operates within a regulated boundary that determines what can be seen and done; interfaces are then not neutral windows but an a priori configuration of constraints and control. Galloway, in a similar vein, theorises the interface as a site of protocol control: a threshold where power acts by governing how information may flow and how interactions may occur (Galloway 2004). Read together, they displace the anthropocentric view of code as instructional text for machines, and show how code is actually enacted through a set of often-invisible software layers as materialised *mediation* between human intentions and technical systems rather than sheer *transmission*, patterned by hardware constraints and protocol governance.

To consider programming in these terms requires attending to the material foundations that make it possible. Taking the Cambridge definition of a program as "a series of instructions that can be put into a computer in order to make it perform an operation," several questions arise: What counts as *instructions*, *computers*, and *operations* in particular configurations? How does one "put" instructions "into" the computer? *At what point* does an *instruction* become an *operation*? These questions point to the boundary-work of transla-tion between the human and the machinic, or what Galloway described as the site "where one significant material is understood as distinct from another significant material," in other words, the interface (2008, 939). It should also be apparent that the interface affording programming is neither settled nor technologically determined; it is instead a set of configurations made concrete under particular ideological and material conditions.

Having clarified the proximity between programming and a study of interfaces, I then discuss its methodological implications. A study of programming interfaces necessarily abstracts: it deconstructs particular interfaces of a system in relation to the users and technologies under inquiry, such that the analysis remains legible in some scope and rationale. Abstraction is then a justified method for scholars to meta-picture the "net" relation between human input and machinic output, allowing analytical movement across "things", "people" and "effects" without collapsing their differences. As Galloway writes, "the interface is never a thing, but processes towards an effect" (2012). This interpretation of an interface is, therefore, relational; one studies relations, not stable objects, as all technical objects are historical. Affordance scholars may find this attitude familiar, where it is common to treat action possibilities as emerging from user-environment couplings, where neither user nor surface remains fixed over time (Gaver 1991; Hutchby 2001; Fayard and Weeks 2014). Noting the unstable and specific nature of human–technology relations, it is important to resist the colloquial temptation to inductively associate certain practices (of interaction) with an a priori conception of the user identity in question, because in such ways we are taking for granted the socially-constructed categorical representations of the human-technical relation. To borrow via Lukács, this would be a form of reification by ossifying a dynamic social relation into a static, natural-seeming "thing," here a user type (Lukács 2021; Fuchs 2021). One cannot assume that issuing a single terminal command defines one as a programmer, nor that a "real" developer must use a particular keyboard, Linux distribution or *Vim*. Analytically, it is the relational configuration through which intentions are articulated and operations are effected *that matters*, alongside its material and ideological construction. From this vantage, programming is situated as a locally stabilised way to translate human intent into machinic effect rather than as an essential feature of computation. A brief historical sketch further demystifies:

The earliest electronic computers, such as ENIAC, demanded direct programming through physical reconfiguration with patch panels and *plugboards* (Haigh et al. 2016). This exemplifies immediate hardware manipulation, where the "program" was materially instantiated as circuit configuration. The stored-program computer introduced symbolic abstraction: machine code and assembly languages enabled human instructions as text—mnemonics for operations, numeric addresses for memory that are stored in memory for execution. Even with machine code, conveying human intentions to machines is conditioned by external authorities (cf. the Intel Management Engine on CPUs) rather than direct transmission-execution. Higher-level languages like *Haskell*, *Lisp*, *Java*, amongst others, delegated interaction to additional software layers including assemblers, compilers, interpreters, virtual machines in exchange for usability. Technologies move along a spectrum of ease and human *legibility* by concealing machinic operations behind stabilised software configurations. Technical knowledge is consequently progressively *re-embedded* into machines as defaults: standard libraries prescribe patterns; IDEs offer reusable snippets; compilers optimise according to *mystic* heuristics; runtimes manage memory and collects garbages. Development ecosystems follow this trajectory, shifting from text editors like *Emacs* and *Vim* to specialised IDEs (Processing, Arduino IDE) to general-purpose environments (VS Code), aggregating scattered components as *stacks* (Bratton 2015). For modern developers, lower-level methods recede, and actions happen through *instrumentalised* metaphors and symbols. In Galloway's terms, such instrumentalisation makes ideology functional, and software becomes the medium through which ideological effects are rendered effective through hardware (Galloway 2012, 69–70).

Human–computer translation as such is never direct or comprehensive. Whether one physically re-patches ENIAC's panels, writes assembly, authors in a word processor, or manipulates GUI objects, the *net effect* remains: a human conveys a *negotiated intention* through actions *afforded* by an *interface*, which in turn produces effects at the hardware level that are eventually rendered legible through system *peripherals* through voltage differences (Kittler 1997, 150). Saying "there is no software" names the extracted, partial status of software as a *category* designed to accommodate an "environment of everyday languages," while the hardware configuration precedes and conditions code in its legible forms. "Software" is best understood as chains of transformation linking inscriptions (source text, compilers, binaries) to operations (electrical states); contemporary programming practice is a particular instance of controlled interaction between human intentions and technical systems mediated by processes of mystification.

Together, Kittler and Galloway deliver a pincer movement for technical media theory. Kittler drills up from the silicon, showing how machine architecture determines what is possible; Galloway analyses down from the screen, revealing how interfaces establish control to produce user effects and ideological occlusions. The consequences for programming are twofold: First, a unprivileged view of programming as but one way to negotiate human intent on computer systems, grounded in infrastructures that make (programming) languages operative. Second, interface analysis must treat control as immanent to mediation: to study an interface is to study how possibilities are structured, constrained, and made accessible.

In the next section I turn to vibe coding as an exemplary case of interface flattening. I first develop a materialist account of vibe coding technologies/practices and survey the discourse around it. I then examine how a model-mediated toolchain establishes a re-layering of human-computer interaction wherein many steps once handled through distinct interfaces (CLI, GUI, API) and symbolic languages are now mediated through a probabilistic intermediary deployed on novel infrastructures. I discuss how this re-layering foregrounds certain action possibilities and resurfaces what it means by a "programming/programmer", while extending particular forms of protocolised control, and how these dynamics make legible the historically specific character of programming languages as interface solutions.

## 3. On the vibe coding paradigm

### 3.1 What *Has Been* Vibe Coding?

Vibe coding emerges as a discourse within computer science and human-computer interaction research primarily focused on empirical workflow analysis and tool development. Recent studies treat vibe coding explicitly as "programming through conversation with code-generating LLMs," focusing on workflows, prompting strategies, and code evaluation (Sarkar et al. 2025; Meske et al. 2025). However, most of these investigations remain confined within the immediate disciplinary of computer science and AI, and treated the phenomenon as a productivity enhancement rather than a *paradigm* shift in computational mediation. As Sarkar and Drosos (2025) note, "vibe coding is an emergent practice that is being continuously defined and negotiated" (4), our analysis focuses on the material and theoretical implications of the paradigm, rather than a pursuit of definition. It is worth noting a parallel strand of literature that engages with the colloquial usage of "vibes" more broadly. Grietzer's "A theory of vibe" (2017) in particular offers a cultural-theoretical account of how "vibes" operate as aesthetic and affective categories in contemporary digital culture. While our focus remains on the technical dimensions of vibe coding, Grietzer's work suggests that the appeal to "vibes" in computational contexts may participate in broader cultural shifts toward *affective* modes of interaction with digital systems.

Drawing from research dicourse on vibe coding (Sarkar and Drosos 2025; Ray 2025; Ross et al. 2023; Brummelen et al. 2020), we can outline the current technical *stack* facilitating this paradigm (**Figure 1**). The *stack* comprises three components: First, an *LLM provider* (3), as a hardware-software configuration that can be local or remote, though in most cases is remotely deployed due to constrains of proprietary softwares (models), computing capacities and power consumption. The *provider* relies on a different hardware architecture char-acterised by GPU clusters and large RAMs than that of typical personal computers, with its operation decoupled from the following components. *Second*, an interface (1) for users to engage in natural language conversation with the LLM. While this can take the basic form of API calls, in practice such interfaces are almost always embedded within broader environments that afford the LLM expanded action possibilities. *Third*, a *toolchain* that mediates between the LLM and computer systems (2), granting it access to various interfaces and existing hardware/software peripherals, such as the ChatGPT application, Cursor IDE, and Claude Code. These technical configurations reveal several material dependencies that distinguish vibe coding from pre-vibe coding programming practices, or how one issues their instructions to a machine: the paradigm requires sustained access to heavy GPU-accelerated computation, whether through local hardware acquisition or cloud service subscription. The computational demands of models also introduce latency and connectivity dependencies in contrast to the immediate responsiveness of a local editing and testing environment. Moreover, the *probabilistic, stochastic* nature of LLM responses introduces a fundamental uncertainty into the translation between human intent and computational operation that was formerly less prevalent with instructions written in programming languages, that reconfigures the development practice to iterative refinement of prompts rather than deterministic, repeatable testing.

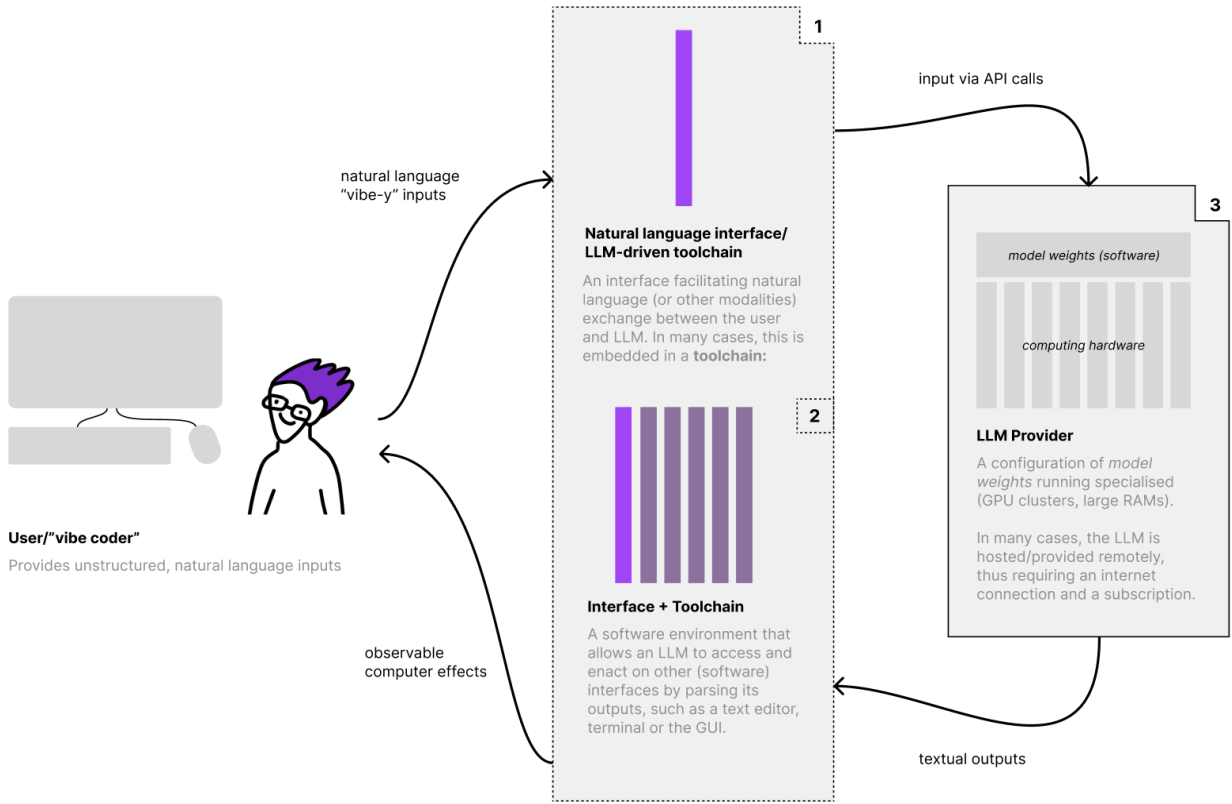

**Figure 1.** The "stack" of vibe coding

## 3.2 Interface Flattening in Action

*Interface flattening*, in the context of vibe coding, names a dynamic by which formerly distinct layers of interaction, whether GUI, CLI or API, dissolve into a single, conversational surface while the underlying chain of translation is quietly extended. It does not replace one interface with another, but routes *intent-actions* through a probabilistic intermediary (a model) that mediates between symbolic utterances and machinic operations via GPU clusters, parsing interfaces, and tool-calling protocols. We arrive at a phenomenology of one *meta*-interface *eclipsing* an infrastructure of "thicker" stacks. Kittler helps specify this flattening/thickening tension with a deconstructive view of "software" as a metaphorical link for computers to co-exist with everyday languages in a chain of translation. Vibe coding did not abolish the chain but added new links for model processing that sit alongside the conventional OS/runtime. The experiential flattening (one prompt box) therefore presupposes an material extension of more layers beneath.

One may confer this shift with the move from "formal" writing to everyday speaking. With vibe coding, the formal interface and its afforded practices of strict syntax, standard development cycles are replaced by unstructured, "vibe-y" inputs on the natural language interface that conceals underlying formal operations. The *symbolic inventory* required to issue the same instructions has changed, resulting in the ostensible "flattening" of the interface felt by the user from intent to observable behaviour. Consider a simple example: prior to vibe coding toolchains, multiple pathways existed for achieving the same material effect of shutting down a computer. One might use the power menu of the operating system's GUI, type in a shell command such as *shutdown*, or, in pure materialist fashion, act on a physi-cal button or cable. A vibe coding approach allows the user to achieve the same effect through natural language instruction within an LLM-driven environment with sufficient system privileges, thereby rendering other interface modalities *secondary* to the user's immediate experience.

Galloway's framework posits the flattened, meta-interface as *protocological.* Even when the prompt affords "vibe-y" interactions, it is governed by schemas for model training, function calling and rate limits. Interfaces are "processes that effect a result" and thresholds where power regulates how actions may flow; vibe coding simply relocates that threshold to the conversation while consolidating control to *model* and *protocol* providers. Tool/function-calling instructions specify model output to typed JSON arguments such that downstream systems could parse and execute; the Model Context Protocol (MCP) standardises how models discover tools and request resources across applications. In this sense, the conversational surface is an *effect* of protocolised coordination where we now delegate models to call a tool, with what arguments to manifest our intent. Ease and flexibility for vibe coders is a success of standardisation *underneath*, suggested by the industry's convergent move towards structured outputs, MCP compliances, and more recently model inter-operability protocols (such as agent2agent by Google).

Keeping in mind the materialities that afforded vibe coding, an LLM *infrastructure* that derives a distribution over likely continuations, and a *toolchain* that parses model output into executable actions, uncovers its implications. *Actions* appear more reachable and *effects* more immediate, at the expense of introducing new dependencies and loci of control: model training and provider policy; power budgets and context windows; state regulations and safety scaffolds. Where pre-vibe coding workflows build upon *relatively static* dependencies of compilers, linters, and libraries that enjoyed an open-source ecosystem, the current configuration defers key decisions to probabilistic inference that is, by design, proprietary and often non-reproducible. Here, Chun's observation (2011) is salient: software nat-uralises ideology by making opaque operations *habitual*. The prompt genre, one that is *affective*, imperative and scattered, encourages users to offload method, memory, and even error-handling into a black-boxed pipeline. Symbolic labour and skills have now shifted to work on the new meta-interface. Imagine now, a media researcher analysing patterns within a corpus of social media texts. Prior to LLM-driven "vibe coding," the researcher would either develop or adapt existing programs for pattern analysis. Now the researcher provides the dataset to an AI/LLM toolchain, describes analytical intentions in plain language, and receives generated scripts or direct results. In many cases, the programming language interface *recedes entirely* as the toolchain executes code in the background (such as ChatGPT's data analysis tool). Yet this apparent seamlessness requires new forms of knowledge: understanding which model to use, data handling limits, and context management—demonstrating how symbolic labour shifts rather than disappears.

This flattening became historically possible because computers can now directly act upon unstructured natural language and increasingly other modalities such as images, sketches, and voices. Following Kittler's understanding of the symbolic-materialist qualities of language and its relative autonomy from nonlinguistic reality, natural language interfaces afforded by LLMs intervene in symbolic transactions with a different kind of *materialisation*, allowing machines themselves to operate within the linguistic. Kittler argued that software would not exist if computer systems did not need to coexist with "an environment of everyday languages" (1997); what we have now is a computer system that *speaks* our everyday languages to an extent far beyond Kittler's times, while exploiting the legible symbolic accommodations afforded by previous software interfaces. Throughout computer development, we have made almost every aspect of the computer operation legible through a close symbolic representation. By doing so, we have paved the way to recursively render the material-technical of the hardware accessible to the symbolic-technical of LLMs, granting ground for a convergence of technical autonomy. In other words, we have handed our symbolic understanding of the machines back to the machines (Weatherby 2025). Interface flattening thus also disrupts the symbolic-linguistic by enabling computers for natural language conversations. The distinction between utterances and their effects is also more ambiguous than ever. Do we now speak a language of the machines, or have they learned to listen? Who and what is the language for? Such questions now beg further investigation.

## 4. Conclusion

By analysing vibe coding as interface flattening, I demonstrated that emergent AI technologies intensify the dynamics that Kittler and Galloway identified as constitutive of human-computer interaction. I argue that vibe coding exemplifies a particular historical moment in which the "chains of transformation" linking human inscription to electrical operation become simultaneously more complex and more concealed despite unshackle from rigid programming languages, and extends the forms of protocol control that govern human-computer interactions. I have shown that *interface flattening* operates through a specific material and ideological arrangement: the convergence of formerly distinct interface modalities into a single conversational surface while the underlying computational infrastructure extends and thickens beneath. This flattening became historically possible because we have made technical operation legible through symbolic representation, and now we trained the symbolic (LLM-as-software) to converge with the historical symbolic scaffolding, thus enabling machines to "operate within the linguistic" in ways that exceed Kittler's original formulation. One may say that we have handed our symbolic understanding of machines back to the machines; its consequences are yet to be known.

My analysis further reveals that apparent democratisation of technical capability in vibe coding operates through the incorporation of previously human activity into technically mediated processes. The *phenomenological ease* of natural language interaction conceals the redistribution of symbolic labour into training processes and infrastructural arrangements that remain inaccessible to direct user engagement. I demonstrate that symbolic labour has shifted to new forms of literacy of prompts, models, and toolchains.

The materialist implications of interface flattening are consequential for understanding the political economy of AI technologies. Vibe coding's dependency upon often-remote large-scale computation and centralised model development represents a reorganisation of the material foundations through which computational capabilities become accessible. This re-organisation effectively privatises the symbolic competencies of (programming) languages and concentrate technical expertise within proprietary systems that were previously distributed across (programming) communities and educational institutions. *Interface flattening reveals*, in retrospective, the historical-specificity of the programming-with-code practice as a locally stabilised arrangement of human-computer interaction. The apparent recession of programming languages within vibe coding workflows demonstrates the evolving forms of computational "control", albeit negotiated and preconfigured, and a requirement for "new" knowledge of honing the meta-interface historically unavailable.

Looking forward, several issues await future investigation. First, the questions of language and its convergence with the machinic: do we now speak a language of the machines, or have they learned to listen? Where do we posit an autonomy of the symbolic *now*? Second, the material foundations of LLM deployment, including environmental costs, labour arrangements and geopolitical agendas require deeper analysis to uncover the broader social implications of LLM-driven interface flattening. Third, investigation into how interface flattening operates across different domains of AI application could reveal whether the patterns I identify in vibe coding represent structural tendencies. Finally, the relationship between interface flattening and the ideological function of LLM-as-softwares reveal the need for understanding how democratisation technologies conceal complex redistributions of agency, control, and dependency within human-technological systems. As AI-mediated interactions shift to a central stage in our digital media landscape, understanding these redistributions becomes essential for developing alternative approaches of computational mediation that *resist* existing forms of centralised control.